\documentclass[12pt]{article}
\input epsf.sty
\newcommand{\rf}{\par\noindent\hangindent 1cm {}}
\begin{document}
\Large
\begin{center}
{\bf
ISO observations of the BL Lac object PKS 2155--304
}
\end{center}

\bigskip
\normalsize
\begin{center}
{\bf 
E.\ Bertone$^1$, G.\ Tagliaferri$^1$, G.\ Ghisellini$^1$, A.\ Treves$^2$, L.\ Chiappetti$^3$, L.\ Maraschi$^1$
}

\bigskip
{\small
{\sl 
$^1$ 
Osservatorio Astronomico di Brera, via Brera 28, I--20121 Milano, Italy
-- email:
bertone@merate.mi.astro.it\\

\smallskip
$^2$ 
Universit\`a dell'Insubria, via Lucini 3, I--22100 Como, Italy \\

\smallskip
$^3$
IFCTR/CNR, via Bassini 15, I--20133 Milano, Italy \\
} 
}
\end{center}


\bigskip
{\bf 
1. Introduction
}

\bigskip\noindent


The Infrared Space Observatory\footnote{ISO is an ESA project with
instruments funded by ESA Member States (especially the PI countries: France,
Germany, the Netherlands and the United Kingdom) with the participation of
ISAS and NASA.} 
(ISO) observed the BL Lac object PKS 2155--304 fifteen times, from May 7 
to June 8 1996. Twelve observations were carried out in
a best sampled period of 15 days. Two
additional observations were performed on 1996 November 23 and 1997 May 15. 
These observations were part of a multiwavelength monitoring which involved
UV, X--ray, $\gamma$--ray satellites and ground based telescopes.

The instruments used were the ISOCAM camera in the range 2.5 -- 18
$\mu$m and the ISOPHOT photometer, from 25 up to 170 $\mu$m.

\bigskip
{\bf 
2. The ISO light curves
}

\begin{figure}[hb]
\vspace{0cm}
\hspace{7.7cm}
\epsfxsize=7.7cm\epsfbox{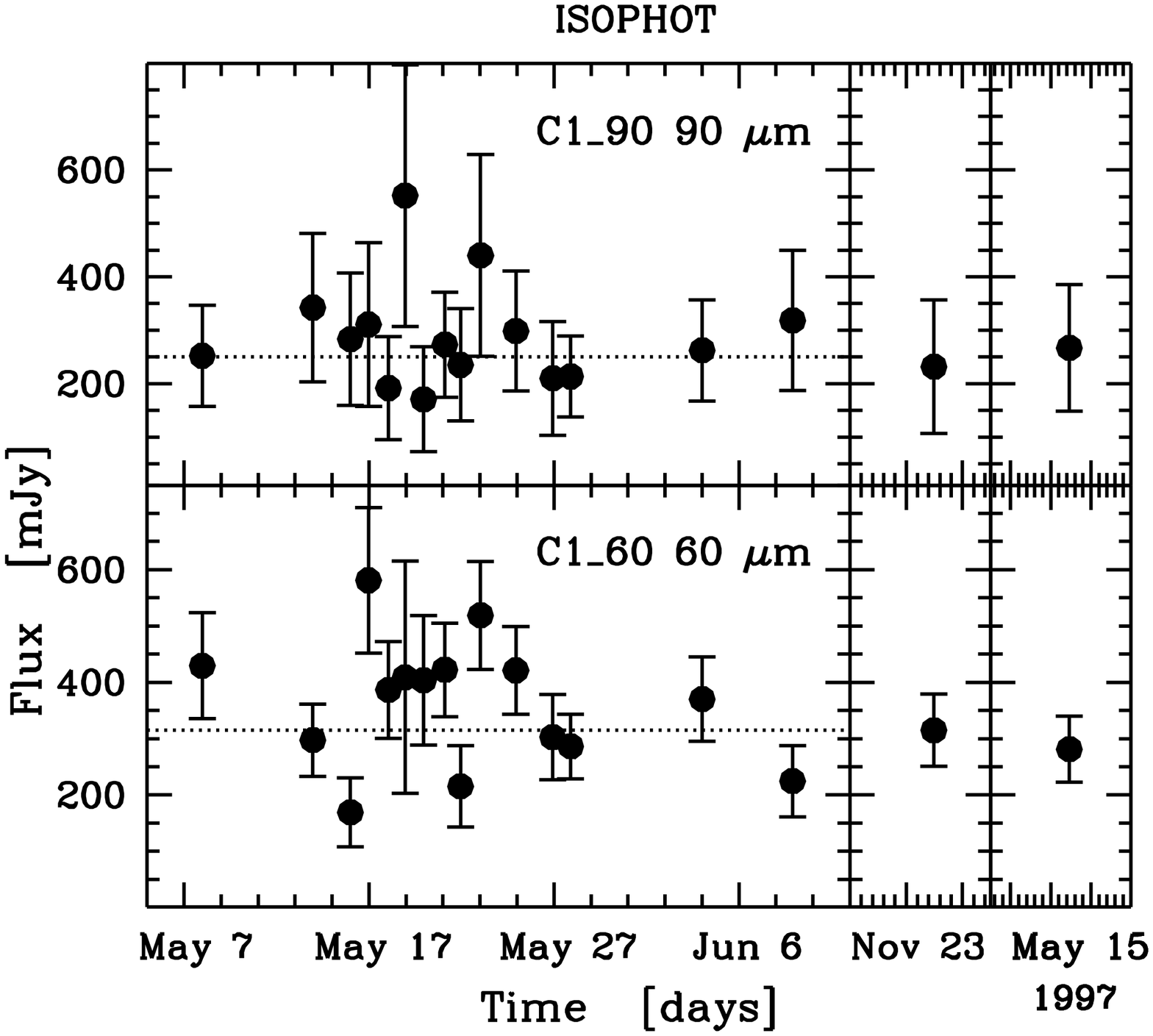}
\vspace{-7.7cm}
\hspace{0cm}\epsfxsize=7.7cm\epsfbox{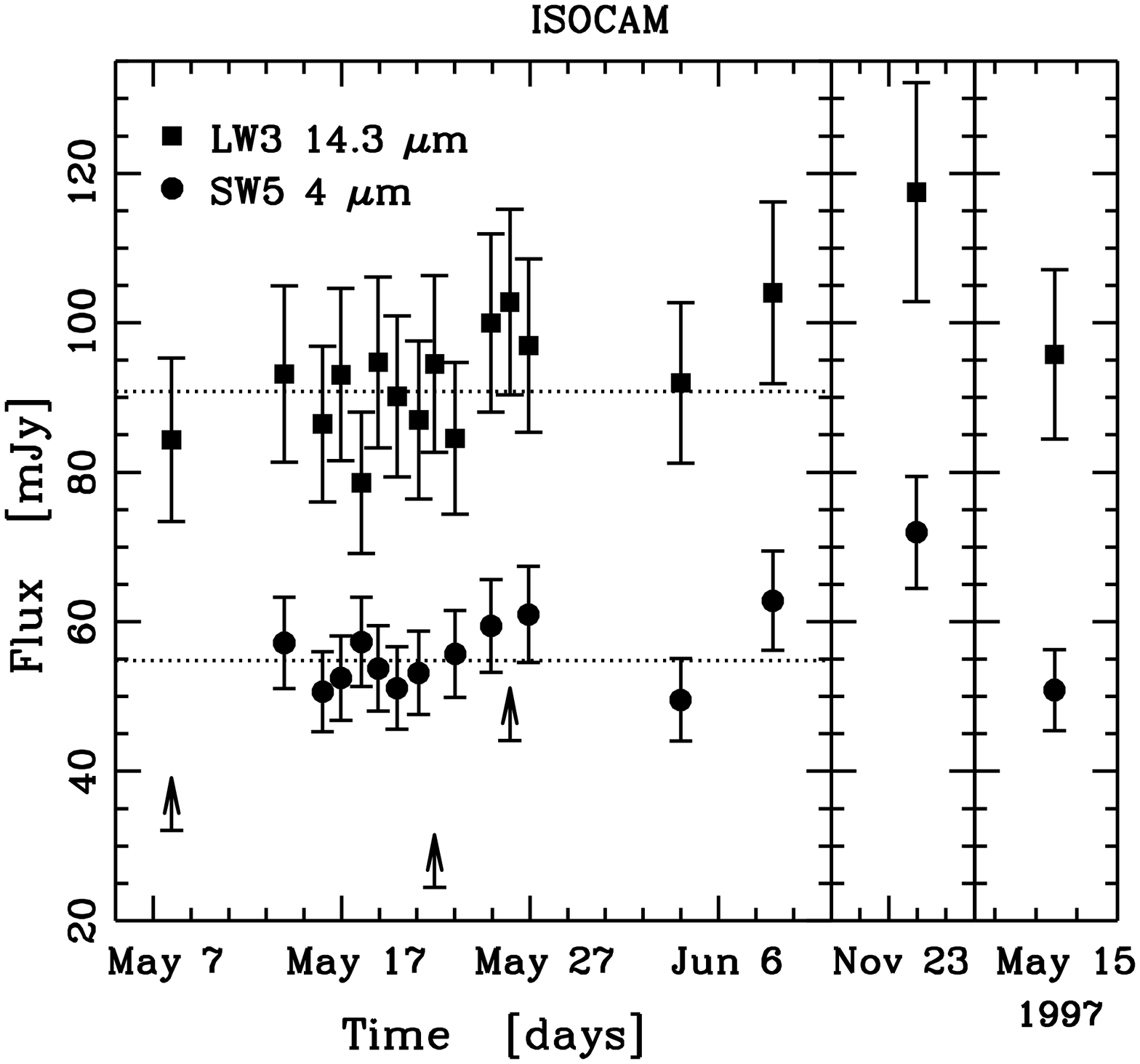}
\caption{\small ISO light curves. Dotted lines are the mean flux values of
the best sampled period. }
\label{fig:lc}
\end{figure}

\bigskip\noindent
The two ISOCAM light curves, at 4.0 and 14.3, $\mu$m and the two ISOPHOT light
curves, at 60 and 90 $\mu$m, are shown in Fig. \ref{fig:lc}. 
There is no evidence of time variability of the flux at the four wavelengths,
but the large errors (more than 10\% for the camera fluxes
and up to 50\% for the photometer data) can hide smaller variations. We
calculated the mean relative error and we obtained 3 sigma limits for the
lowest detectable flux variations of 32\%, 36\%, 76\%  and 132\% at 4.0, 14.3,
60 and 90 $\mu$m respectively.
The mean flux values of the best sampled period, from 1996 May 13 to
May 27, are  $54.8 \pm 1.8$ mJy at 4.0 $\mu$m, $90.8 \pm 3.2$ mJy 
at 14.3 $\mu$m, $315 \pm 27$ mJy at 60 $\mu$m and $250 \pm 34$ mJy at 
90 $\mu$m.

\bigskip
{\bf 
3. The ISO spectrum
}

\begin{figure}[t]
\vspace{0cm}
\hspace{7.7cm}
\epsfxsize=7.7cm\epsfbox{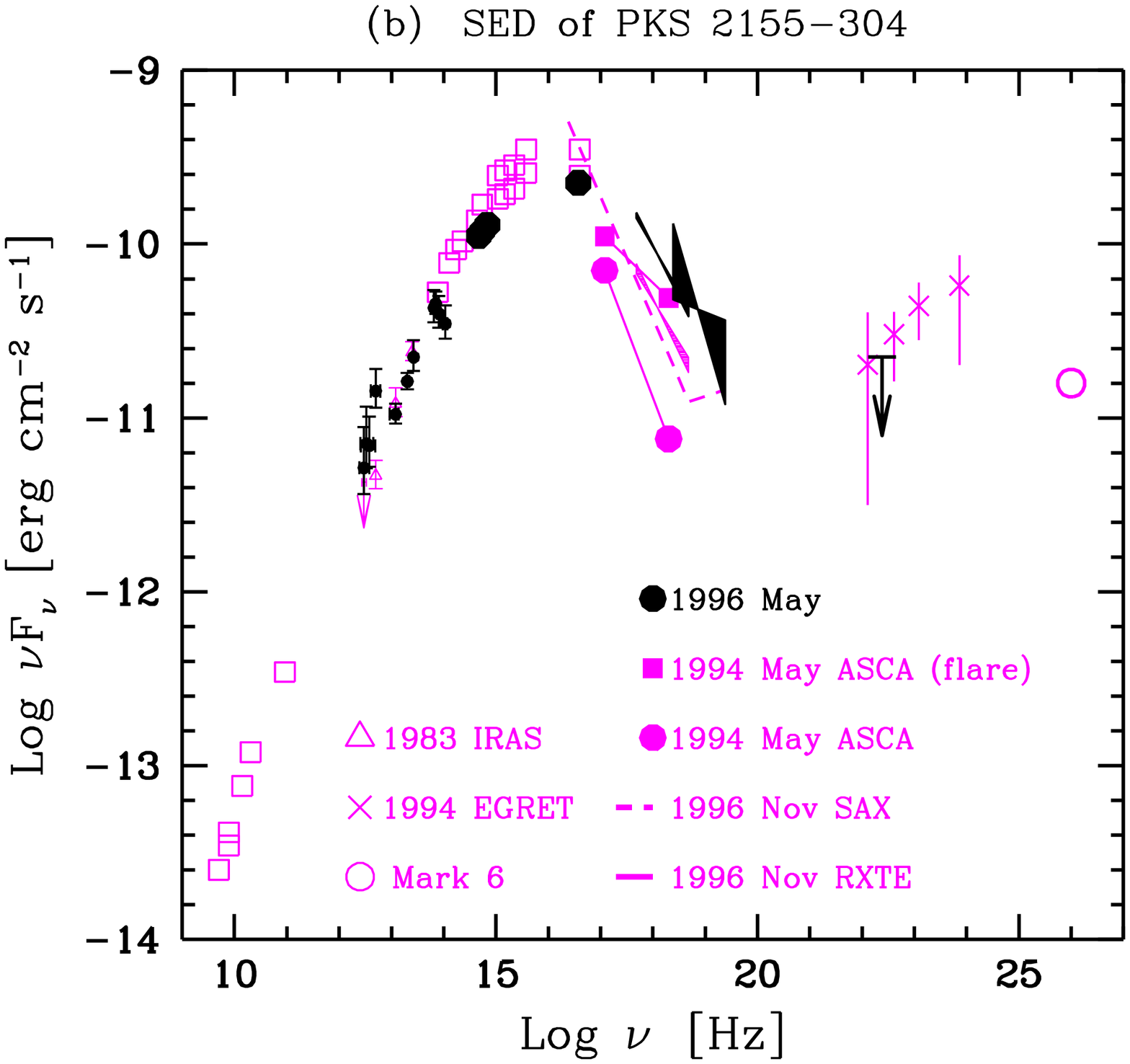}
\vspace{-7.7cm}
\hspace{0cm}
\epsfxsize=7.7cm\epsfbox{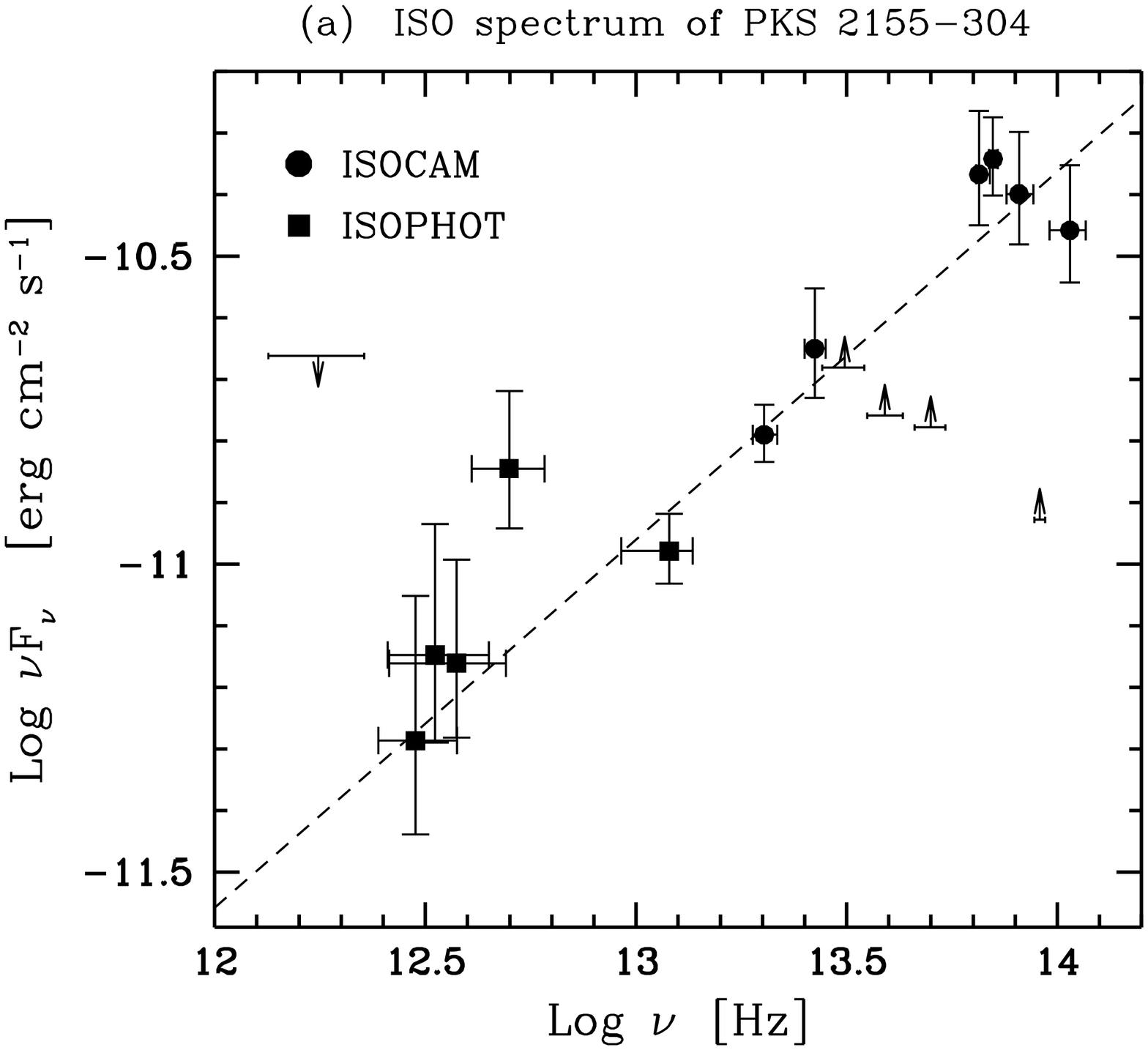}
\vspace{0cm}
\caption{\small {\bf (a)} ISO spectrum. {\bf (b)} Spectral energy distribution of PKS 2155--304. Black points are
the simultaneous data of 1996 May (Bertone et al. 1999; Urry et al. 1998;
Marshall \ H.L., priv. comm.; Vestrand \ W.T., priv. comm.). Grey data are
from literature.}
\label{fig:spectrumsed}
\end{figure}

\bigskip\noindent
The infrared spectrum of PKS 2155--304 was sampled, using 16 broad band
filters, from 2.8 to 170 $\mu$m (Fig. \ref{fig:spectrumsed}a). It is well
fitted with a single power law of energy spectral index $\alpha = 0.40 \pm 0.06$.

This power--law shape strongly supports the hypothesis that the far-- and
mid-- infrared emission of this source is entirely generated by synchrotron process, excluding, therefore, important contributions from thermal sources. 
The emission of the host galaxy of PKS 2155--304, a big elliptical that is seen
in NIR images (Kotilainen et al. 1998), is negligible at longer IR wavelengths
(Bertone et al. 1999).

From the SED of the source in Fig. \ref{fig:spectrumsed}b, which
shows the simultaneous data of 1996 May, one can notice that the ISO data lie
well on the extrapolation between the radio and the optical data, and follow
well the shape of the synchrotron peak, supporting the fact that the
synchrotron emission is dominant in the ISO band.

\bigskip
{\bf References}

\bigskip
{\small
\rf
Bertone E., Tagliaferri G., Ghisellini G., et al., 1999, in preparation
\rf
Kotilainen J.K., Falomo R. \& Scarpa R., 1998, A\&A 336, 479
\rf
Urry C.M., Sambruna R.M., Brinkmann W.P. \& Marshall H., 1998, in: Scarsi L.,
Bradt H., Giommi P., Fiore F.\ (eds.) The Active X-Ray Sky. Nucl.\ Phys.\ B (Proc.\ Suppl.) 69, 419 
}
\end{document}